# Analysis of laser polarization state on remote induced plasma luminescence characteristics of filament in air

Yuezheng Wang [1,2], Zhi Zhang [1,3], Zeliang Zhang [1,2], Nan Zhang [1,2], Lie Lin [1,3], Weiwei Liu [1,2*]

[1] Institute of Modern Optics, Eye Institute, Nankai University, Tianjin 300350, China;
[2] Tianjin Key Laboratory of Micro-scale Optical Information Science and Technology, Tianjin 300350, China;
[3] Tianjin Key Laboratory of Optoelectronic Sensor and Sensing Network Technology, Tianjin 300350, China

**Abstract**: The femtosecond laser filamentation is the result of the dynamic interplay between plasma self-focusing and defocusing generated by the multiphoton/tunnel ionization of air molecules. This equilibrium allows the filament to stably propagate over long distances at high power densities, making it a promising tool for remote sensing in chemical and biological applications, detection of air pollution, and lightning control, which has attracted wide attention. Laser-induced filamentation is a highly nonlinear process, and initial conditions such as changes in polarization state can affect the fluorescence emission of $N_2$ molecules excited by the filament in air. In this article, we analyze in detail the effect of polarization state changes on the 337 nm fluorescence signal excited by the filament at a distance of 30 m. It was found that the fluorescence signal intensity of the linear polarization state was higher than that of the circular polarization state. Through analysis of these phenomena, we explore the mechanism of plasma fluorescence emission during the long-range filamentation process of femtosecond laser, including the electron collision model and the polarization effect on the critical power for filamentation. These findings are important for the understanding of the stimulated radiation from filaments and may find applications in remote sensing of electric field and THz radiation.

**Keywords**: femtosecond laser filamentation; Plasma luminescence; Laser polarization

## 1. Introduction

When a high-power femtosecond laser pulse is transmitted through a transparent medium, such as air or glass, it can overcome the natural diffraction effect and achieve long-distance transmission of 0-10 km by maintaining a dynamic balance between the plasma defocusing effect and the Kerr self-focusing effect[1-5]. In this scenario, laser pulses are concentrated into a cylindrical channel with a diameter of approximately 100 μm[6, 7], known as the "filament". Since Braun and his colleagues first reported the filamentation of femtosecond laser pulses in air in 1995, it has garnered significant attention due to its broad potential applications in various fields[8, 9], such as intense terahertz emission[10-12], harmonic generation[13], artificial rainfall[14, 15], remote sensing of atmospheric composition[16-18], laser processing[19], and air laser generation, among others[20-22].

The light intensity inside the femtosecond laser filament in air can reach $5 \times 10^{13}$ W/cm$^2$ (intensity clamping)[23, 24], part of the molecules and atoms are ionized, and part of the small molecular fragments enter the excited state, both of which can emit clean fluorescence[25, 26]. Measuring the spectrum along the filament propagation path enables us to extract not only the plasma density, electron temperature, and laser intensity within the filament, but also to gain insight into the excitation and ionization processes that occur during filament formation[27-30]. More importantly, remote sensing of atmospheric pollutants can be achieved by detecting the backward scattering characteristic fingerprint luminescence left by the plasma generated by femtosecond pulses during filament formation in air[18, 31]. However, laser-induced filamentation is a highly nonlinear process, which makes it highly sensitive to experimental conditions such as external focusing conditions[32], initial beam radius[33], laser pulse duration, and polarization[34, 35]. This inevitably has a strong impact on various phenomena related to laser filamentation. For example, recent studies have shown that fluorescence and laser generated in the filament are strongly influenced by the polarization state of the driving laser.

Liu et al. used a commercial laser system to study the effect of laser polarization on the excitation of nitrogen fluorescence signals by filamented light at a distance of 1 meter. They found that at lower laser pump energies (250 μJ), linearly polarized laser excites stronger nitrogen fluorescence signals than circularly polarized laser, but the opposite phenomenon occurs at higher laser pump energies (8.3 mJ). They attributed this phenomenon to the higher optical field ionization rate of linearly polarized laser at lower pump energy, while the circularly polarized laser pulse above the threshold intensity generated a new channel of impact excitation by high-energy electrons[36]. At a distance of 40 cm, Mingxing Jin and his team experimentally measured the angular distribution of plasma emission during the filamentation process of a linearly polarized femtosecond laser. They observed that the $N_2^+$ fluorescence emission was stronger in the direction parallel to the polarization of the laser, and the signal intensity at 337 nm increased linearly with pressure. They proposed a direct intersystem crossing scheme to explain this phenomenon[37]. Up to now, there has not been a detailed report on how the laser polarization state at longer filament distances affects the nitrogen fluorescence signal excited by filaments.

In this paper, we investigate the emission fluorescence of nitrogen induced by filamentation of femtosecond laser with long focal length (30m). We focus on observing the changes in the intensity of radiation fluorescence from the second positive band system ($C^3\Pi_u^+ - B^3\Pi_g^+$ transition; 337–nm line) of nitrogen under various polarization states. This not only contributes to the application of laser filaments in remote sensing of atmospheric composition and air laser, but also provides new data for the investigation of the mechanism of plasma fluorescence emission during filament formation.

## 2. Experimental Details

The experiments were performed with a Ti:Sapphire femtosecond laser system (Spectra Physics), as shown in Figure 1. The system delivers pulses of 50 fs at 800 nm with a maximum single pulse energy of 4 mJ, a repetition frequency of 500 Hz, and a beam diameter of approximately 13 mm ($1/e^2$ level of intensity). The femtosecond laser filament is formed at a distance of 30 meters through a reflection off-axis system, and its length is approximately 45 cm. The filament formation system uses a combination lens, and the distance of filament formation can be precisely controlled by adjusting the relative distance between the concave lens (F = -15 cm; R = 2.54 cm) and the concave mirror (F = 200 cm; R = 20.7 cm). In addition, a quarter-wave plate is placed in front of the combined lens to change the laser polarization direction of the initial line polarization pulse. The generated plasma fluorescence is first collected by the lens (F=50 mm) fixed on the propagation path mobile station, and then enters the monochromator (WGD-100, Gang Dong Sci. & Tech. Co. Ltd.), a photomultiplier tube (PMT, Hamamatsu, H11902) and a programmable timing generator (DG535, Stanford Research Systems) to obtain the fluorescence spectrum signal of nitrogen (337nm) in the air environment excited by the filament. The entrance slit of the monochromator was arranged to be parallel to the plasma filament in order to increase the fluorescence collection efficiency. In addition, to improve the signal-to-noise ratio, each data point in this paper is typically the average accumulation of 10 sets of 128 shots.

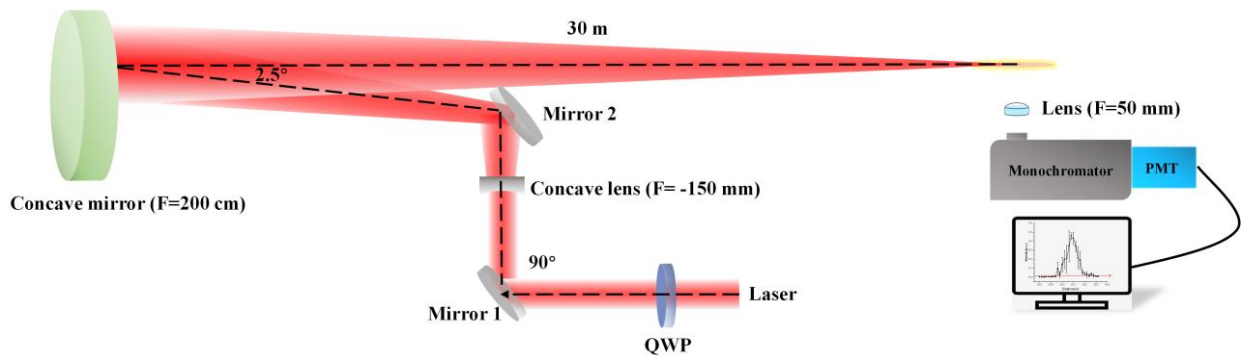

**Fig. 1** Schematic diagram of the device for femtosecond laser filamentation to excite nitrogen fluorescence. QWP: quarter-wave plate; Mirror 1 and Mirror 2：Broadband Dielectric Mirrors；Lens：UV Fused Silica Plano-Convex Lenses, AR Coated: 245 - 400 nm.

In our experiments, we tried to study the effect of the polarization of the laser pulse on its excitation of nitrogen molecules to produce fluorescence. The polarization state of the laser pulse is changed by rotating the quarter waveplate, when the rotation angle is 45° corresponding to left circular polarization state, 135° corresponding to right circular polarization state, 0° and

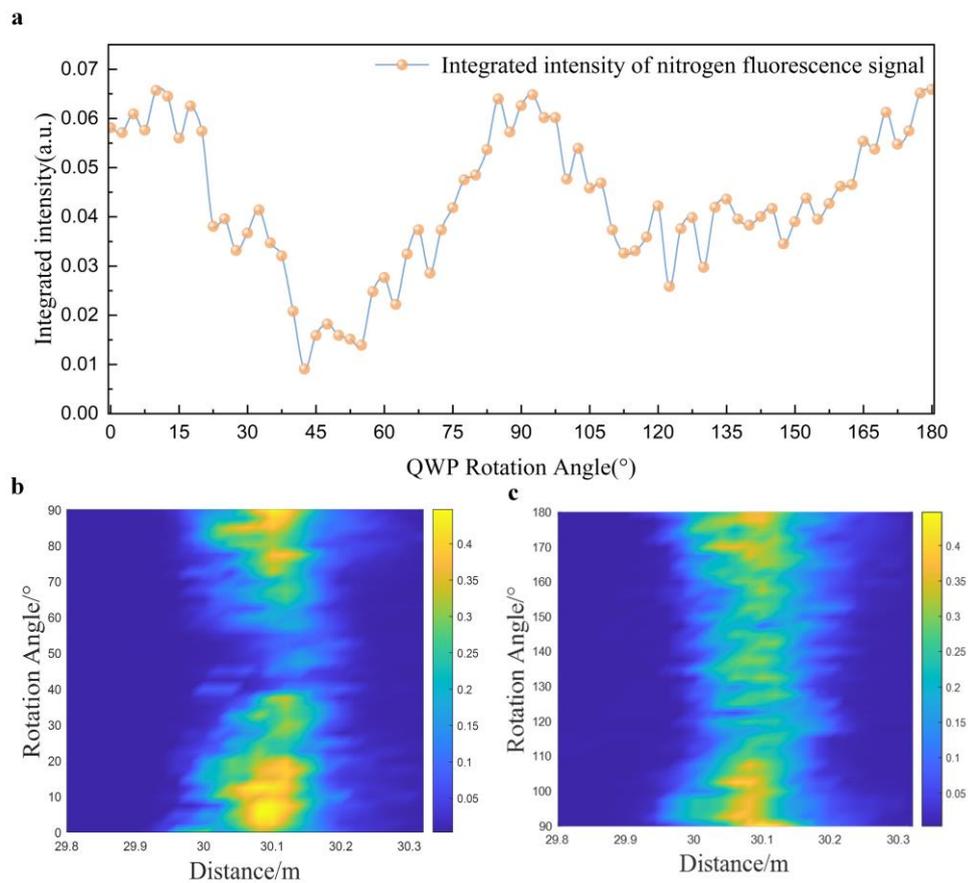

**Fig. 2 a** Variation of the integrated intensity value of the fluorescence signal (337 nm) of a filament-excited nitrogen molecule with the laser polarization state (0°/180°- horizontal linear polarization, 45°- left circular polarization state, 135°- right circular polarization state, 90°- vertical linear polarization state). **b** Nitrogen fluorescence signal integrated intensity pseudo-color map (0~90°). **c** Nitrogen fluorescence signal integrated intensity pseudo-color map (90~180°).

180° corresponding to horizontal linear polarization state, 90° corresponding to vertical linear polarization state. In addition, the remaining rotation angle corresponds to the elliptical polarization state.

## 3. Results and Discussion

Fig. 2 shows the evolution of the 337 nm spectral line intensity along the propagation axis as the air is irradiated by the circularly, elliptically and linearly polarized pulses with same energies. When the equivalent focal length of the combined lens is 30 m, the signal intensity of 337 nm in the line polarization case is much higher than that in the circular polarization case, and the signal intensity decreases rapidly as the ellipticity decreases. For this result, the analysis is that the fluorescence at 337 nm originates from the $C^3\Pi_u(v'=0) \rightarrow B^3\Pi_g(v''=0)$ of N$_2$. Its intensity is proportional to the amount of N$_2$(C$^3\Pi_u$). According to the electron collisional excitation model[38], the excited state of nitrogen (C$^3\Pi_u$) is produced by the excitation of a neutral nitrogen molecule by a high-energy electron impact, which can be expressed as follows:

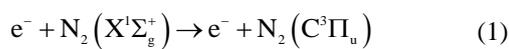
$$e^- + N_2\left(X^1\Sigma_g^+\right) \rightarrow e^- + N_2\left(C^3\Pi_u\right) \quad (1)$$

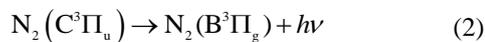
$$N_2\left(C^3\Pi_u\right) \rightarrow N_2(B^3\Pi_g) + h\nu \quad (2)$$

The basic ionization process of a molecule or atom under strong laser irradiation can be divided into three steps[39]. First the electrons tunnel through the potential barrier formed by the Coulomb potential and the laser field, then the tunneling electrons are accelerated by the laser field, and finally some electrons are driven back and then collide with their parent ions. From the above equations 1 and 2, it can be seen that the filament excitation nitrogen fluorescence intensity in air is proportional to the electron collision excitation rate. Also, according to the law of conservation of charge, the yield of ions is equal to the yield of electrons. Since the tunneling ionization rate is exponentially related to the instantaneous electric field of the laser, the reduction of the maximum electric field will greatly suppress the ionization rate when the laser deflection state is changed from linear polarization to circular polarization. When the laser intensity is low, the recombination probability between electrons and their parent ions is higher in the linearly polarized state than in the circularly polarized state, which also results in a stronger fluorescence signal at 337 nm in the linearly polarized state. Our experimental results demonstrate that filamentation of femtosecond laser induces nitrogen molecular fluorescence (337 nm) in air at long distances, which is better explained by the electron collision model. Specifically, the excited states of nitrogen molecules($C^3\Pi_u$) are generated through collision between high-energy electrons and neutral nitrogen molecules. This finding shed new light on the mechanism of filamentation-induced fluorescence in air.

In the experiment of exciting nitrogen fluorescence using femtosecond laser filamentation at close range, researchers generally believe that the energy window for electron collisional excitation is between 12 and 30 eV[40]. Within this energy window, the electron yield of the linearly polarized laser field is much higher than that of the circularly polarized laser field. As the laser energy increases, the relative intensity of the nitrogen fluorescence excited by linearly polarized and circularly polarized lasers reverses. For example, in the experiment conducted by Mingxing Jin et al., when the laser energy exceeded the "energy threshold" (2.0 mJ) at a distance of 1m, the nitrogen 337nm fluorescence emission under circular polarization was stronger[37]. Similarly, in the

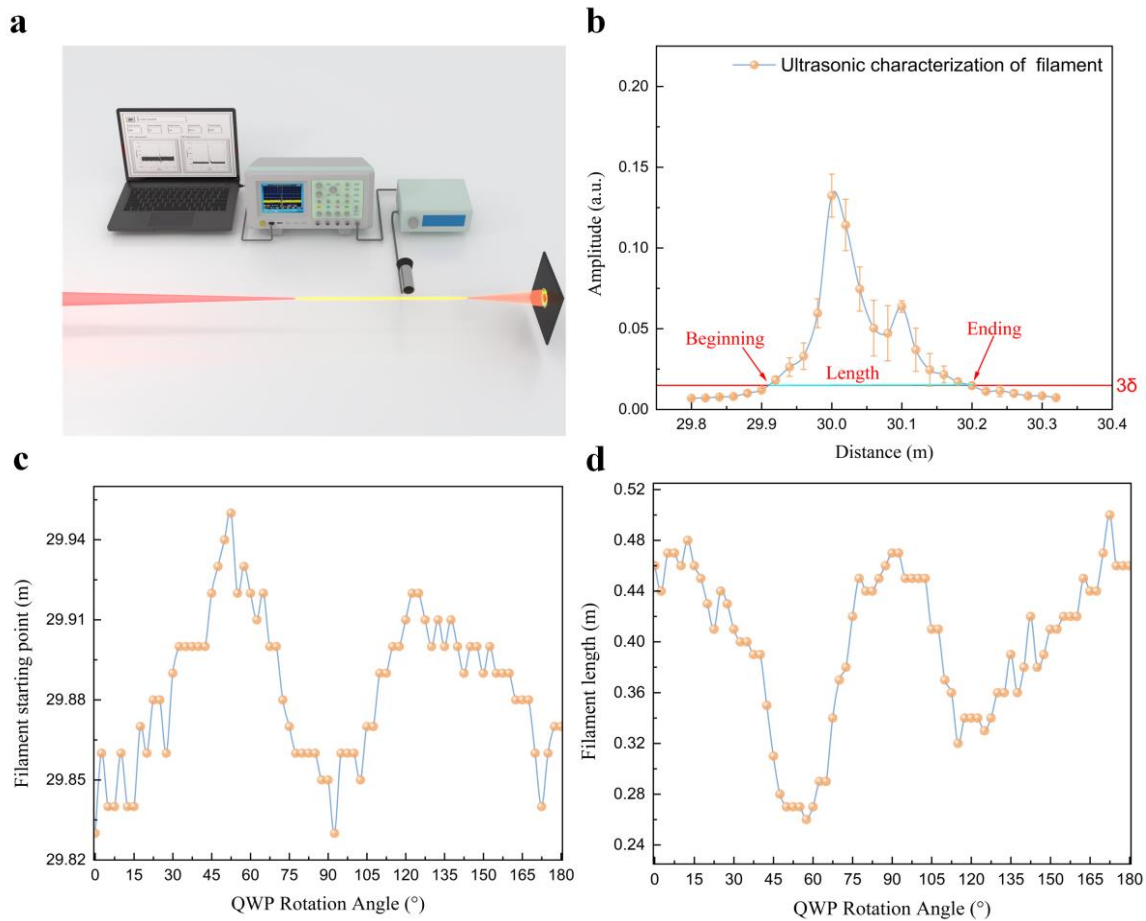

**Fig. 3 a** Schematic diagram of the femtosecond laser filamentation setup characterized by ultrasonic signals. **b** Extracting filament starting point and length information from ultrasonic signal image example (the red line represents

3 times the standard deviation of the ambient noise 3σ). **c** Variation of the filament starting point in different polarization states (QWP Rotation Angle：0°/180°- horizontal linear polarization,45°- left circular polarization state, 135°- right circular polarization state, 90°- vertical linear polarization state). **d** Variation of filament length in different polarization states.

experiment conducted by Sergey Mitryukovskiy et al. at a distance of 1m, it was found that the nitrogen 337nm fluorescence emission under linear polarization was higher than that under circular polarization when the laser energy was 250 μJ, while under circular polarization, the nitrogen 337nm fluorescence emission was stronger when the laser energy was 8.3 mJ.

However, in the experiment of filamentation at long distance, there was no observed phenomenon of reversal of the relative intensity of fluorescence emission under linearly polarized and circularly polarized laser excitation. Our hypothesis is that at long distances, the polarization state of the laser affects the critical power for self-focusing in air, thereby changing the second-order nonlinear refractive index $n_2$ of air. Circularly polarized light has a higher critical power for self-focusing than linearly polarized light, resulting in a more likely filamentation process in the linear polarization state, thereby causing nitrogen at 337 nm to emit stronger fluorescence under the electron collision model.

To verify this hypothesis, an experiment as shown in Figure 3a was designed to characterize the variation of the starting point and length of femtosecond laser filamentation under different polarization states. The methods commonly used to characterize filamentation include electromagnetic pulse (EMP) detection, $N_2/N_2^+$ fluorescence detection in the wavelength range of 300-420 nm, and ultrasonic detection[41]. The principle of characterizing filament using ultrasonic signals is as follows: During the process of photoionization, free electrons are ejected with high kinetic energy on the order of several eV, and their initial electron temperature is around $10^4 \sim 10^5$ K. The energy transfer between the hot free electron gas and heavy electrons in the surrounding gas is accomplished through collisions of the thermal gas column produced by plasma recombination. The expansion of this thermal gas column generates acoustic wave (AW) emission, which can be detected through a microphone, and the intensity of the sound wave is proportional to the laser energy in the filament. By moving the microphone along the filament, the spatial range and longitudinal plasma distribution of the filament can be determined[42].

In this paper, we placed an ultrasonic probe (V306, Olympus) on one side of the filament to collect the acoustic signal excited by the filament. The acoustic signal was amplified by an ultrasonic pulse receiver (5072PR, Olympus) and displayed on a digital oscilloscope (DPO3034, Tektronix), which was recorded by a computer. The starting point, length, and axial relative intensity distribution of the femtosecond laser filament in this experiment were recorded in detail by using a manual displacement table to move the ultrasonic probe (MHz) along the axis of the filament with a step of 2 cm. After simple data processing, we can obtain the spatial information of the complete filament in a certain polarization state, as shown in Figure 3b. By comparing the axial relative intensity with three times the standard deviation of the noise, we can read out the starting position and length of the filament from the Figure 3b. Next, by rotating a quarter-wave plate with a step size of 2.5° (range: 0~180°), we obtained the variation of the starting point and the length of the filament with the polarization state, as shown in Figure 3c and Figure 3d, respectively. It can be seen that under linear polarization, the starting point of the filament moves forward, and the filament length is the longest. As the ellipticity increases, the starting point of the filament gradually moves backward, and the filament length gradually becomes shorter. The phenomenon of the focal position moving towards the lens (forward shift) can be explained by the joint contribution of external focusing and self-focusing to the input laser beam, as given by the following equation (3) and equation (4)[43]:

$$1/f' = 1/f + 1/z_f \qquad (3)$$

$$P_{cr} = \frac{3.77\lambda^2}{8\pi n_2 n_0} \qquad (4)$$

In the equation, $f$ represents the focal length of the external focusing lens, $z_f$ represents the self-focusing distance of the collimated Gaussian beam, and $P_{cr}$ represents the critical power for self-focusing. When the laser pulse exceeds the critical power and further increases, the focal spot of the laser beam will move towards the external focusing lens. Linearly polarized light has a lower critical power, and under the same pulse laser energy, it is more prone to the phenomenon of focal shift. Meanwhile, the increase in the length of the filament leads to an increase in plasma density, which results in a higher probability of electron collisions in the electron collision model, consistent with our hypothesis.

**4. Conclusion**

This work investigates the influence of laser polarization on the plasma emission fluorescence in air during femtosecond laser filamentation over a long distance (30m). It is found that under linear polarization, nitrogen emits stronger fluorescence at 337nm, whereas under circular polarization, it is weaker. During this process, the intensity of the fluorescence signal decreases as the ellipticity increases. The explanation for this phenomenon can be divided into two aspects. Firstly, the mechanism of nitrogen excited state ($C^3\Pi_u$) emitting fluorescence at 337nm induced by filamentation is believed to be the free electron collision model. When the laser polarization state changes from linear to circular, the reduction of the maximum electric field greatly suppresses the ionization rate and reduces the probability of secondary collision between electrons and their parent ions, resulting in a decrease in fluorescence intensity. On the other hand, linear polarization light has a lower critical power for self-focusing than circular polarization light, which makes the pulse laser more prone to filamentation under linear polarization, thereby leading to stronger nitrogen fluorescence at 337 nm. Our research results provide more clues for understanding the luminescence mechanism of $N_2$ molecules during air filamentation.

## References

1. A. Braun, G. Korn, X. Liu, D. Du, J. Squier, and G. Mourou, "Self-channeling of high-peak-power femtosecond laser pulses in air," Opt Lett **20**, 73-75 (1995).
2. R. Y. Chiao, E. Garmire, and C. H. Townes, "Self-Trapping of Optical Beams," Physical Review Letters **13**, 479-482 (1964).
3. S. Xu, J. Bernhardt, M. Sharifi, W. Liu, and S. L. Chin, "Intensity clamping during laser filamentation by TW level femtosecond laser in air and argon," Laser Physics **22**, 195-202 (2012).
4. M. Rodriguez, R. Bourayou, G. Mejean, J. Kasparian, J. Yu, E. Salmon, A. Scholz, B. Stecklum, J. Eisloffel, U. Laux, A. P. Hatzes, R. Sauerbrey, L. Woste, and J. P. Wolf, "Kilometer-range nonlinear propagation of femtosecond laser pulses," Phys. Rev. E **69**, 7 (2004).
5. I. Dicaire, V. Jukna, C. Praz, C. Milian, L. Summerer, and A. Couairon, "Spaceborne laser filamentation for atmospheric remote sensing," Laser Photon. Rev. **10**, 481-493 (2016).
6. J. Kasparian, M. Rodriguez, G. Mejean, J. Yu, E. Salmon, H. Wille, R. Bourayou, S. Frey, Y. B. Andre, A. Mysyrowicz, R. Sauerbrey, J. P. Wolf, and L. Woste, "White-light filaments for atmospheric analysis," Science **301**, 61-64 (2003).
7. A. Couairon, and A. Mysyrowicz, "Femtosecond filamentation in transparent media," Phys. Rep.-Rev. Sec. Phys. Lett. **441**, 47-189 (2007).
8. J. Kasparian, and J. P. Wolf, "Physics and applications of atmospheric nonlinear optics and filamentation," Opt Express **16**, 466-493 (2008).
9. P. Qi, W. Qian, L. Guo, J. Xue, N. Zhang, Y. Wang, Z. Zhang, Z. Zhang, L. Lin, C. Sun, L. Zhu, and W. Liu, "Sensing with Femtosecond Laser Filamentation," Sensors **22**, 7076 (2022).

10. T. I. Oh, Y. J. Yoo, Y. S. You, and K. Y. Kim, "Generation of strong terahertz fields exceeding 8 MV/cm at 1 kHz and real-time beam profiling," Applied Physics Letters **105** (2014).
11. L.-L. Zhang, W.-M. Wang, T. Wu, S.-J. Feng, K. Kang, C.-L. Zhang, Y. Zhang, Y.-T. Li, Z.-M. Sheng, and X.-C. Zhang, "Strong Terahertz Radiation from a Liquid-Water Line," Physical Review Applied **12**, 014005 (2019).
12. A. D. Koulouklidis, C. Gollner, V. Shumakova, V. Y. Fedorov, A. Pugzlys, A. Baltuska, and S. Tzortzakis, "Observation of extremely efficient terahertz generation from mid-infrared two-color laser filaments," Nat. Commun. **11**, 8 (2020).
13. C. Vozzi, F. Calegari, E. Benedetti, S. Gasilov, G. Sansone, G. Cerullo, M. Nisoli, S. De Silvestri, and S. Stagira, "Millijoule-level phase-stabilized few-optical-cycle infrared parametric source," Opt. Lett. **32**, 2957-2959 (2007).
14. J. J. Ju, J. S. Liu, C. Wang, H. Y. Sun, W. T. Wang, X. C. Ge, C. Li, S. L. Chin, R. X. Li, and Z. Z. Xu, "Laser-filamentation-induced condensation and snow formation in a cloud chamber," Opt. Lett. **37**, 1214-1216 (2012).
15. J. P. Wolf, "Short-pulse lasers for weather control," Rep. Prog. Phys. **81**, 34 (2018).
16. H. L. Xu, and S. L. Chin, "Femtosecond Laser Filamentation for Atmospheric Sensing," Sensors **11**, 32-53 (2011).
17. J. Kasparian, and J. P. Wolf, "Physics and applications of atmospheric nonlinear optics and filamentation," Opt. Express **16**, 466-493 (2008).
18. S. L. Chin, H. L. Xu, Q. Luo, F. Theberge, W. Liu, J. F. Daigle, Y. Kamali, P. T. Simard, J. Bernhardt, S. A. Hosseini, M. Sharifi, G. Mejean, A. Azarm, C. Marceau, O. Kosareva, V. P. Kandidov, N. Akozbek, A. Becker, G. Roy, P. Mathieu, J. R. Simard, M. Chateauneuf, and J. Dubois, "Filamentation "remote" sensing of chemical and biological agents/pollutants using only one femtosecond laser source," Appl. Phys. B-Lasers Opt. **95**, 1-12 (2009).
19. X. Zhan, H. Xu, C. Li, H. Zang, C. Liu, J. Zhao, and H. Sun, "Remote and rapid micromachining of broadband low-reflectivity black silicon surfaces by femtosecond laser filaments," Opt Lett **42**, 510-513 (2017).
20. Q. Luo, W. Liu, and S. L. Chin, "Lasing action in air induced by ultra-fast laser filamentation," Appl. Phys. B-Lasers Opt. **76**, 337-340 (2003).
21. J. P. Yao, B. Zeng, H. L. Xu, G. H. Li, W. Chu, J. L. Ni, H. S. Zhang, S. L. Chin, Y. Cheng, and Z. Z. Xu, "High-brightness switchable multiwavelength remote laser in air," Phys. Rev. A **84**, 5 (2011).
22. J. P. Yao, W. Chu, Z. X. Liu, J. M. Chen, B. Xu, and Y. Cheng, "An anatomy of strong-field ionization-induced air lasing," Appl. Phys. B-Lasers Opt. **124**, 17 (2018).
23. J. Kasparian, R. Sauerbrey, and S. L. Chin, "The critical laser intensity of self-guided light filaments in air," Appl. Phys. B-Lasers Opt. **71**, 877-879 (2000).
24. W. Liu, S. Petit, A. Becker, N. Akozbek, C. M. Bowden, and S. L. Chin, "Intensity clamping of a femtosecond laser pulse in condensed matter," Opt. Commun. **202**, 189-197 (2002).
25. A. Becker, A. D. Bandrauk, and S. L. Chin, "S-matrix analysis of non-resonant multiphoton ionisation of inner-valence electrons of the nitrogen molecule," Chem. Phys. Lett. **343**, 345-350 (2001).
26. A. Talebpour, M. Abdel-Fattah, A. D. Bandrauk, and S. L. Chin, "Spectroscopy of the gases interacting with intense femtosecond laser pulses," Laser Physics **11**, 68-76 (2001).
27. S. Q. Xu, X. D. Sun, B. Zeng, W. Chu, J. Y. Zhao, W. W. Liu, Y. Cheng, Z. Z. Xu, and S. L. Chin, "Simple method of measuring laser peak intensity inside femtosecond laser filament in air," Opt. Express **20**, 299-307 (2012).
28. S. A. Hosseini, Q. Luo, B. Ferland, W. Liu, N. Akozbek, G. Roy, and S. L. Chin, "Effective length of filaments measurement using backscattered fluorescence from nitrogen molecules," Appl. Phys. B-Lasers Opt. **77**, 697-702 (2003).
29. X. D. Sun, S. Q. Xu, J. Y. Zhao, W. W. Liu, Y. Cheng, Z. Z. Xu, S. L. Chin, and G. G. Mu, "Impressive laser intensity increase at the trailing stage of femtosecond laser filamentation in air," Opt. Express **20**, 4790-4795 (2012).
30. J. F. Daigle, A. Jaron-Becker, S. Hosseini, T. J. Wang, Y. Kamali, G. Roy, A. Becker, and S. L. Chin, "Intensity clamping measurement of laser filaments in air at 400 and 800 nm," Phys. Rev. A **82**, 5 (2010).
31. R. Bourayou, G. Mejean, J. Kasparian, M. Rodriguez, E. Salmon, J. Yu, H. Lehmann, B. Stecklum, U. Laux, J. Eisloffel, A. Scholz, A. P. Hatzes, R. Sauerbrey, L. Woste, and J. P. Wolf, "White-light filaments for multiparameter analysis of cloud microphysics," J. Opt. Soc. Am. B-Opt. Phys. **22**, 369-377 (2005).

32. F. Theberge, W. W. Liu, P. T. Simard, A. Becker, and S. L. Chin, "Plasma density inside a femtosecond laser filament in air: Strong dependence on external focusing," Phys. Rev. E **74**, 7 (2006).

33. Q. Luo, S. A. Hosseini, W. Liu, J. F. Gravel, O. G. Kosareva, N. A. Panov, N. Akozbek, V. P. Kandidov, G. Roy, and S. L. Chin, "Effect of beam diameter on the propagation of intense femtosecond laser pulses," Appl. Phys. B-Lasers Opt. **80**, 35-38 (2005).

34. S. Rostami, M. Chini, K. Lim, J. P. Palastro, M. Durand, J. C. Diels, L. Arissian, M. Baudelet, and M. Richardson, "Dramatic enhancement of supercontinuum generation in elliptically-polarized laser filaments," Sci Rep **6**, 8 (2016).

35. H. S. Zhang, C. R. Jing, G. H. Li, H. Q. Xie, J. P. Yao, B. Zeng, W. Chu, J. L. Ni, H. L. Xu, and Y. Cheng, "Abnormal dependence of strong-field-ionization-induced nitrogen lasing on polarization ellipticity of the driving field," Phys. Rev. A **88**, 5 (2013).

36. S. Mitryukovskiy, Y. Liu, P. Ding, A. Houard, A. Couairon, and A. Mysyrowicz, "Plasma Luminescence from Femtosecond Filaments in Air: Evidence for Impact Excitation with Circularly Polarized Light Pulses," Physical Review Letters **114**, 5 (2015).

37. Y. Shi, A. M. Chen, Y. F. Jiang, S. Y. Li, and M. X. Jin, "Influence of laser polarization on plasma fluorescence emission during the femtosecond filamentation in air," Opt. Commun. **367**, 174-180 (2016).

38. W. Zheng, Z. M. Miao, C. Dai, Y. Wang, Y. Liu, Q. H. Gong, and C. Y. Wu, "Formation Mechanism of Excited Neutral Nitrogen Molecules Pumped by Intense Femtosecond Laser Pulses," J. Phys. Chem. Lett. **11**, 7702-7708 (2020).

39. Corkum, "Plasma perspective on strong field multiphoton ionization," Physical review letters **71**, 1994-1997 (1993).

40. Y. Itikawa, "Cross sections for electron collisions with nitrogen molecules," J. Phys. Chem. Ref. Data **35**, 31-53 (2006).

41. S. A. Hosseini, J. Yu, Q. Luo, and S. L. Chin, "Multi-parameter characterization of the longitudinal plasma profile of a filament: a comparative study," Appl. Phys. B-Lasers Opt. **79**, 519-523 (2004).

42. J. Yu, D. Mondelain, J. Kasparian, E. Salmon, S. Geffroy, C. Favre, V. Boutou, and J.-P. Wolf, "Sonographic probing of laser filaments in air," Appl. Opt. **42**, 7117-7120 (2003).

43. W. Liu, and S. L. Chin, "Direct measurement of the critical power of femtosecond Ti:sapphire laser pulse in air," Opt. Express **13**, 5750-5755 (2005).

Acknowledgements

This research was funded by the National Key Research and Development Program of China (2018YFB0504400) and Fundamental Research Funds for the Central Universities (63223052).

Author contributions

Yuezheng Wang: Data curation, Writing – original draft, Software. Zhi Zhang: Conceptualization, Methodology, review & editing. Zeliang Zhang: Data curation. Lie Lin: Resources. Weiwei Liu: Conceptualization, Methodology, Validation, Supervision.

Competing interests

The authors declare no competing financial interests.